\documentclass{PoS}

\def\be{\begin{equation}}
\def\ee{\end{equation}}
\def\bea{\begin{eqnarray}}
\def\eea{\end{eqnarray}}

\def\tev{\, {\rm TeV}}
\def\gev{\, {\rm GeV}}

\newcommand{\gsim}{\lower.7ex\hbox{$\;\stackrel{\textstyle>}{\sim}\;$}}
\newcommand{\lsim}{\lower.7ex\hbox{$\;\stackrel{\textstyle<}{\sim}\;$}}

\newcommand{\ifb}{\rm fb^{-1}}

\newcommand{\Dsle}[1]{\slash\hskip -0.20 cm #1}
\newcommand{\met}{{\Dsle E_T}}

\title{Large Jet Multiplicities and Natural Supersymmetry at the LHC}

\ShortTitle{Large Jet Multiplicities and Natural SUSY at the LHC}

\author{\speaker{Jason Kumar}\\
        Department of Physics and Astronomy, University of
Hawai'i, Honolulu, HI 96822, USA\\
        E-mail: \email{jkumar@hawaii.edu}}

\abstract{We consider a search strategy for new physics at the Large Hadron
Collider which focuses on the signature of many jets and missing transverse
energy, but no charged leptons.  We show that this signature can be useful
in probing a wide class of models, including natural supersymmetry,
in which dark matter is produced in conjunction top quarks.  As an example,
we apply this strategy to a simplified supersymmetric model with a light
gluino, light stop and light neutralino.
The efficacy of this strategy is comparable to (and in some
cases better than) that of other strategies which require charged leptons and/or $b$-tagged jets.}

\FullConference{36th International Conference on High Energy Physics,\\
		July 4-11, 2012\\
		Melbourne, Australia}

\begin{document}

\section{Introduction}

There are several scenarios for new physics containing heavy QCD-charged
particles which are also charged under a new unbroken discrete
symmetry and which couple mostly to 3rd generation
Standard Model matter~\cite{ModelsWithJetsPlusMET}.  The new heavy
particles can be produced copiously at the Large Hadron Collider (LHC), and
will necessarily decay to top/bottom quarks (due to $SU(3)$ gauge-invariance) and to
the lightest particle charged under the discrete symmetry (which is a
dark matter candidate).

An example which has drawn recent interest is {\it natural supersymmetry}~\cite{NaturalSUSY},
in which a top squark is relatively light, while other squarks are heavy.
Natural supersymmetry is one way of reconciling relatively small fine-tuning with
LHC constraints on the mass of 1st generation squarks.  A
characteristic feature of natural supersymmetry would be the pair-production of
color-charged particles through QCD processes, with each new particle decaying to
top quarks and missing transverse energy.  It is thus
vital to consider methods of probing these models with LHC data.

We consider a search strategy which selects events with a large number of jets (not necessarily $b$-tagged)
and missing transverse energy, but no charged leptons~\cite{Bramante:2011xd}
(see also~\cite{JetsPlusMET,JetsPluMETExp}).
We will find that our search strategy is advantageous for this scenario of new physics.
In particular, the charged lepton veto will remove Standard Model
events where the missing transverse energy arises from $W$ production followed by the leptonic
decay $W \rightarrow l \nu$.  After this veto, the dominant Standard Model background
satisfying the jet and missing transverse energy cuts is top pair-production;
since the background also contains $b$-jets, $b$-tagging is unnecessary.
The are two main kinematic regimes, in which the energy of the new pair-produced particles
emerges either with the visible particles, or as~$\met$.  In the latter case, an
elevated~$\met$-cut is useful.

\section{An Example Supersymmetric Model}

To implement our search strategy, we will consider an example model
in which there is minimal flavor violation and
the only light particles of the MSSM spectrum are the
lightest neutralino ($\tilde N_1$), the lightest stop ($\tilde t_1$)
and the gluino ($\tilde g$).  In this case, the dominant sparticle
production processes at the LHC are $pp \rightarrow \tilde g \tilde g$
and $pp \rightarrow \tilde t_1^* \tilde t_1$.  Since all of couplings
in the Feynman diagrams for these processes are determined by $SU(3)$
gauge-invariance, both production cross-sections are determined
by the masses of the lightest stop ($m_{\tilde t_1}$) and the gluino
($M_{\tilde g}$).  We will assume $m_{\tilde t_1} > m_t + M_{\tilde N_1}$,
$M_{\tilde g} > 2m_t + M_{\tilde N_1}$.

Similarly, the sparticle decay chains are entirely determined by
gauge-invariance, R-parity conservation, and minimal flavor violation.  The
allowed decay chains are $\tilde t_1 \rightarrow t \tilde N_1$ and
$\tilde g \rightarrow \bar t \tilde t_1 \rightarrow \bar t t \tilde N_1$, where
the decay rates are again entirely determined by the masses of the sparticles.
Each hadronic top decay will produce three partons.  Stop pair-production will
nominally result in six jets, while gluino pair-production will
nominally result in twelve jets.\footnote{In conventional parlance, twelve jets
constitutes a squadron.}

For any point in our low-energy parameter space, a complete supersymmetric model is generated
using SuSpect 2.41~\cite{SuSpect}.  Signal and background events are generated
with MadGraph/MadEvent 5 and Pythia~\cite{Madgraph,PYTHIA}, and  PGS~\cite{pgs} is used
to simulate detector effects (see~\cite{Bramante:2011xd} for details).  NLO corrections to the signal production cross-section
are computed using Prospino~\cite{Prospino}, while NLO corrections to background
cross-sections are estimated from results in the literature~\cite{NLObgd}.

\section{Event Selection}

We begin with an initial set of cuts applied to all selected events
\begin{itemize}
\item{No isolated $e^\pm$ or $\mu^\pm$ in the final state, and
at least five isolated jets ($p_T > 40~\gev$ for each jet, $\Delta R_{jj} > 0.4$ for each pair of jets).}
\item{At least $11.5^\circ$ angular separation between the missing momentum
direction and the 3 leading jets (this cut serves to remove QCD background events
where missing transverse energy arises from jet mismeasurement~\cite{CutsKillQCDBG}).}
\item{$\met > 100~\gev$}
\end{itemize}
The leading contribution to Standard Model events passing these cuts are from
$W,Z+$jets and $\bar t t$ (with a mistagged lepton or a hadronically decaying
$\tau$).  The $\bar t t Z$ background is negligible.

\begin{figure}[tb]
\includegraphics*[width=0.30\columnwidth]{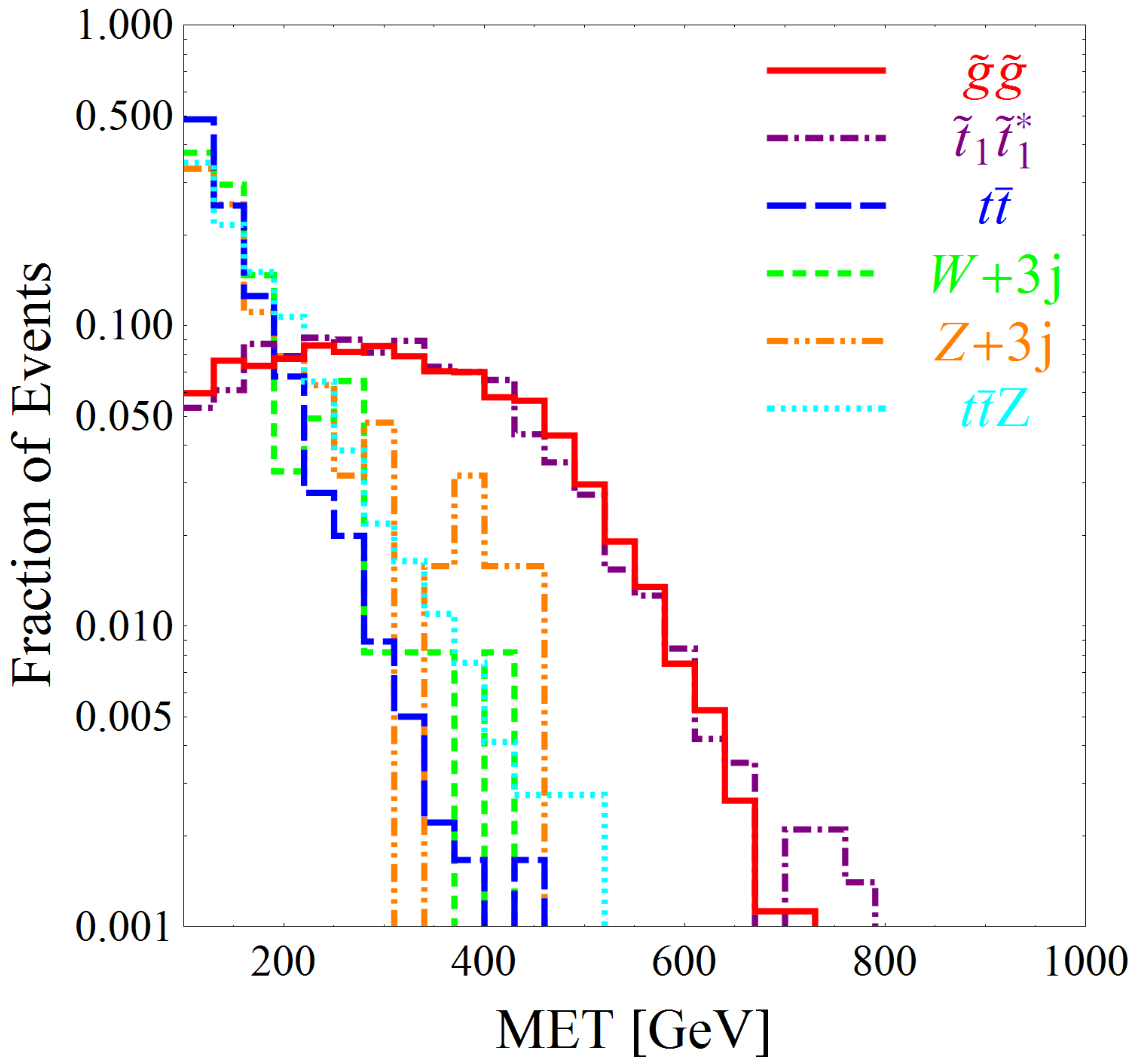}
\includegraphics*[width=0.30\columnwidth]{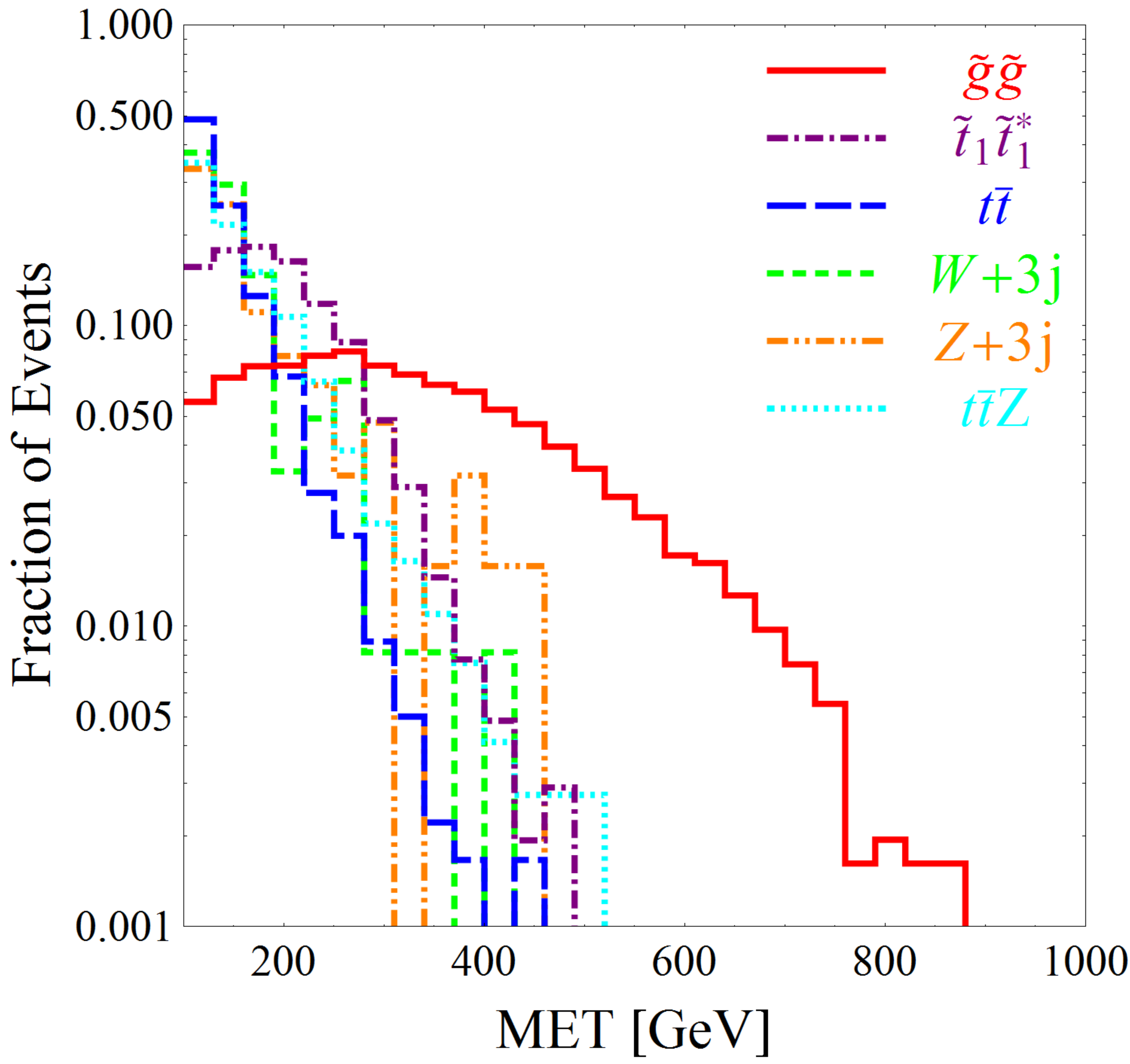}
\includegraphics*[width=0.30\columnwidth]{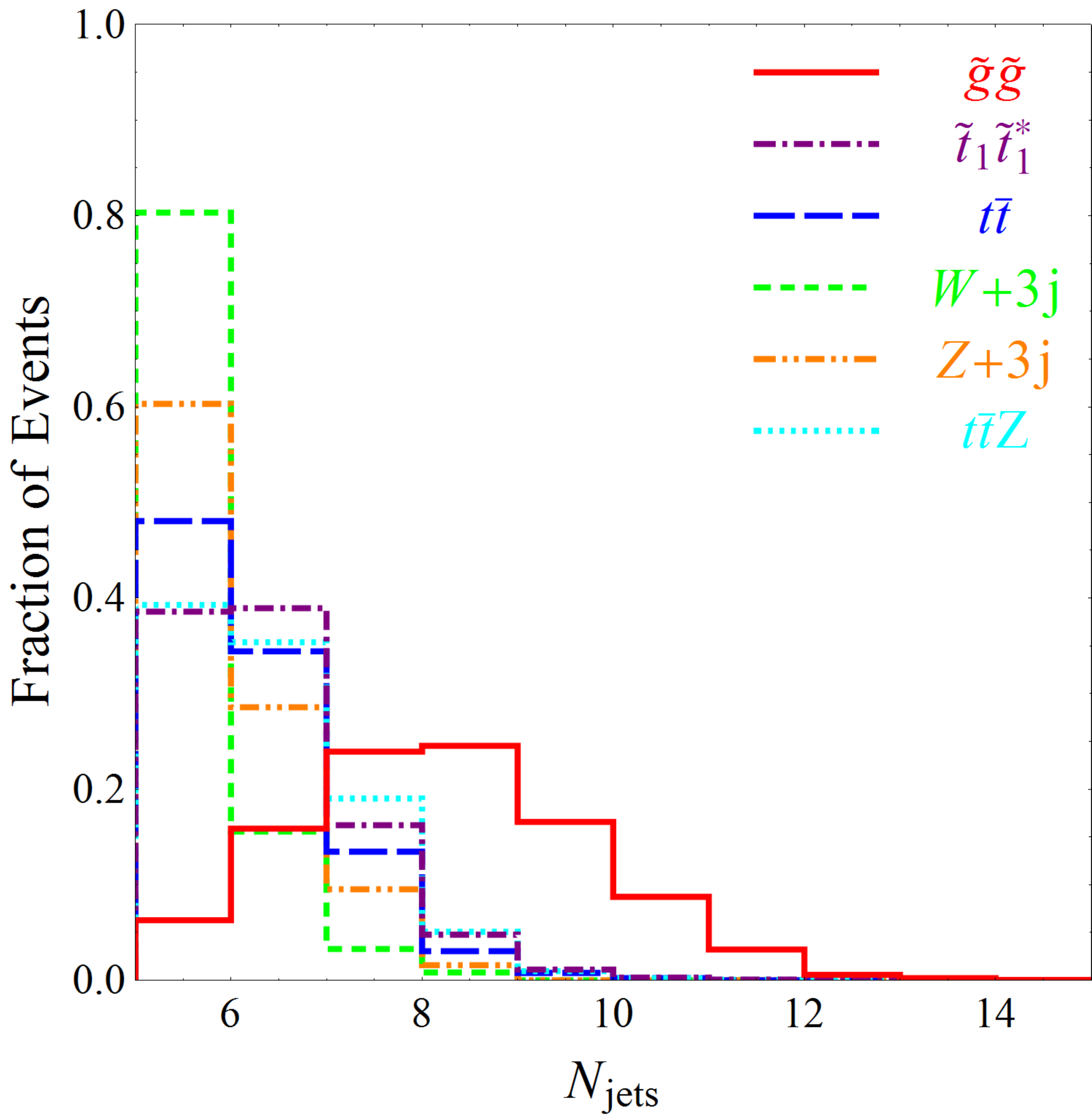}
\vspace*{-.1in}
\caption{\label{fig:METDist}
$\met$-distribution (left and center) and jet distribution (right) for background
and signal events (as labeled) passing the precuts.  We assume $M_{\tilde N_1}=100~\gev$, and either
$M_{\tilde g}=800~\gev$, $m_{\tilde t_1}=600~\gev$ (left),
$M_{\tilde g}=1200~\gev$, $m_{\tilde t_1}=400~\gev$ (center) or
$M_{\tilde g}=1000~\gev$, $m_{\tilde t_1}=600~\gev$ (right).}
\end{figure}

In figure~\ref{fig:METDist} we plot the $\met$-distribution for signal and background
events which satisfy these precuts.  We assume $M_{\tilde N_1}=100~\gev$ and either
$M_{\tilde g}=800~\gev$, $m_{\tilde t_1}=600~\gev$ (left panel) or
$M_{\tilde g}=1200~\gev$, $m_{\tilde t_1}=400~\gev$ (center panel).  We also
plot the distribution of isolated jets for signal and background (right panel)
in the case $M_{\tilde g}=1000~\gev$, $m_{\tilde t_1}=600~\gev$, $M_{\tilde N_1}=100~\gev$.
These distributions illustrate the key features which can be used in distinguishing
signal from background.  As the right panel shows, one expects many more jets from
gluino pair-production than stop pair-production; gluino production is a prototype for
models wherein the decay of new particles produces more than two $t$'s, and an elevated
jet cut can improve signal significance.

The major remaining uncertainty lies in the kinematics of the heavy particle decay, which
can yield energy mostly in the tops (and thus, the visible jets),
or in~$\met$.  The case $M_{\tilde g}=800~\gev$, $m_{\tilde t_1}=600~\gev$ provides
one example; the decay of the gluino produces a stop which is nearly at rest, but
the stop decay produces an energetic neutralino.  The case $M_{\tilde g}=1200~\gev$,
$m_{\tilde t_1}=400~\gev$ provides a different example; gluino production and subsequent
decay produces a boosted stop and leads to a boosted neutralino, but stop production and
subsequent decay produces a top and a neutralino nearly at rest.

Larger mass splittings between
particles in the decay chain generally lead to larger $\met$, but the particular $\met$-distribution
is model-dependent.  In addition to the precuts, it can thus be useful
to elevate the cut on the number of jets (to search for models producing more than two tops)
and/or elevate the $\met$ cut (to search for models whose decay chains produce boosted dark matter).

\section{Discovery Potential }

The discovery potential of this search strategy is shown in figure~\ref{fig:SignificancePanels}.
Here we show signal significance contours in the $(m_{\tilde t_1}, M_{\tilde g})$ plane
with $10~\ifb$ of 7 TeV LHC data (we again assume $M_{\tilde N_1}=100~\gev$).  The orange and
red contours refer to $3\sigma$ and $5\sigma$ Gaussian equivalent significance, respectively.  In
the dark gray hashed region, there are fewer than 5 signal events.  In the upper left panel, only the
precuts are imposed.  In the upper right panel, and additional $\met > 300~\gev$ cut is imposed.
In the lower left panel, and additional cut on the number of jets, $N_j \geq 8$ is imposed.  Finally,
in the lower right panel, both additional cuts $\met > 300~\gev$ and $N_j \geq 8$ are imposed.

\begin{figure}[ht!]
\centerline{
  \includegraphics*[width=0.45\columnwidth]{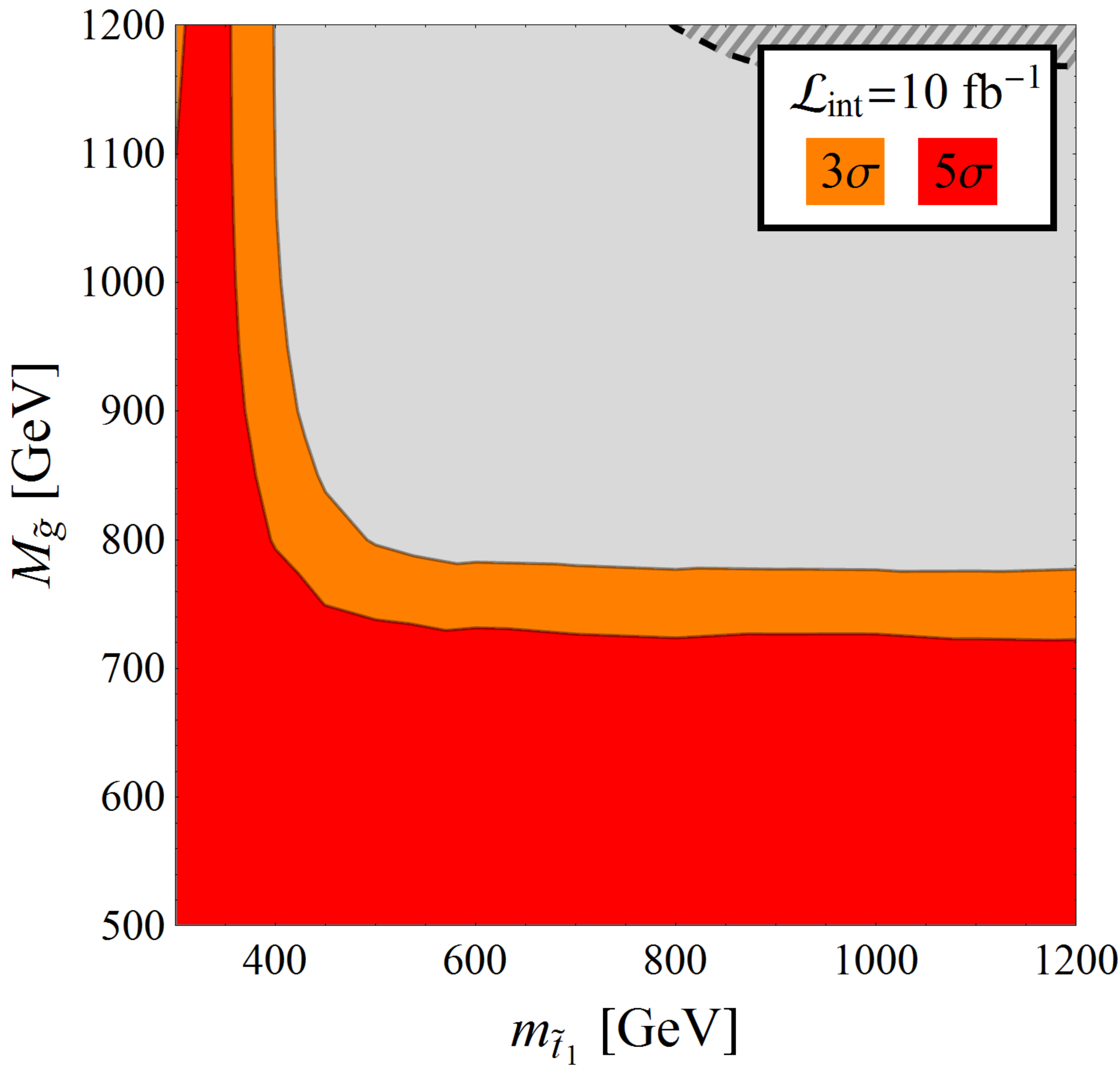}
  \includegraphics*[width=0.45\columnwidth]{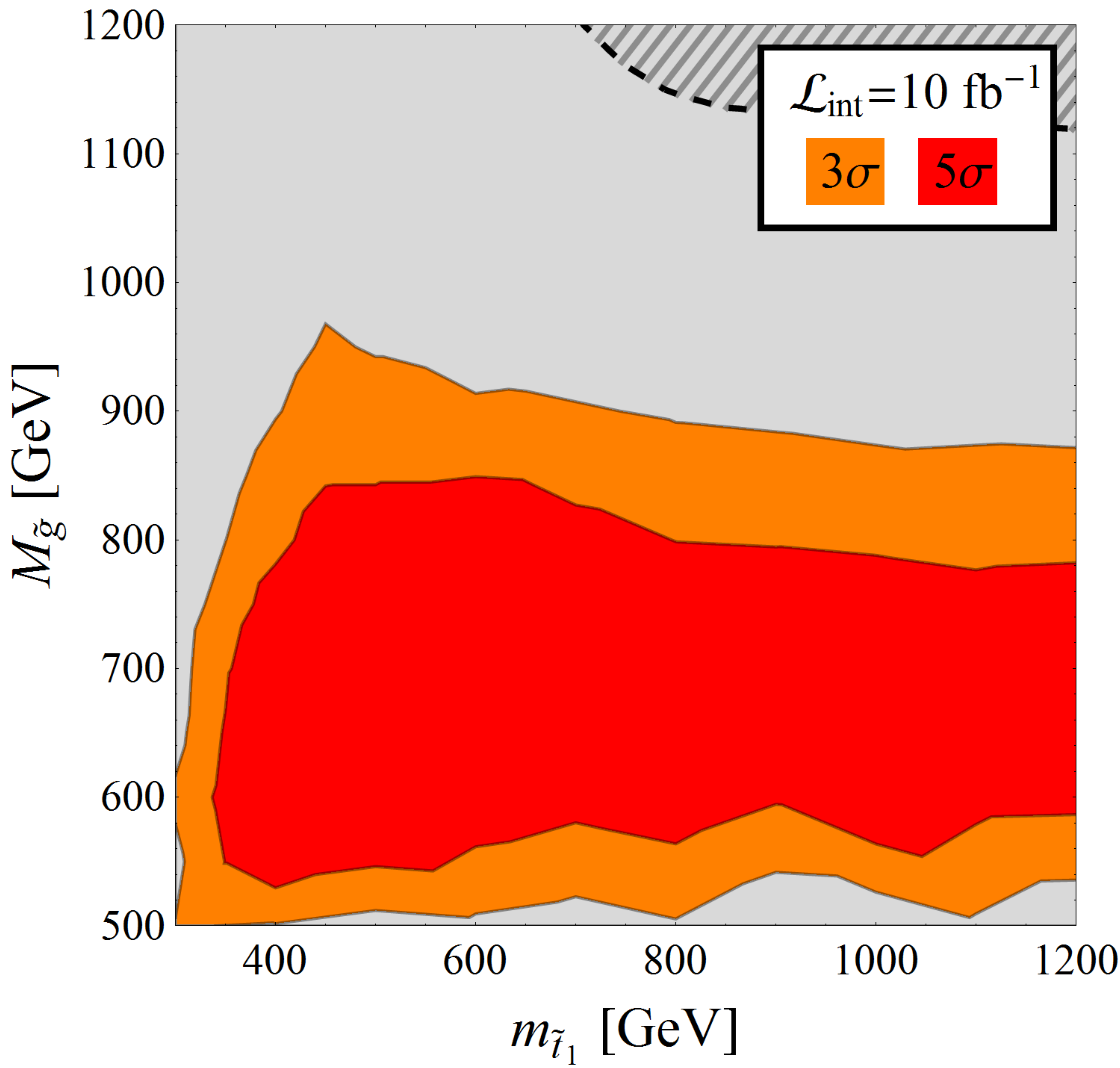} }
\centerline{
  \includegraphics*[width=0.45\columnwidth]{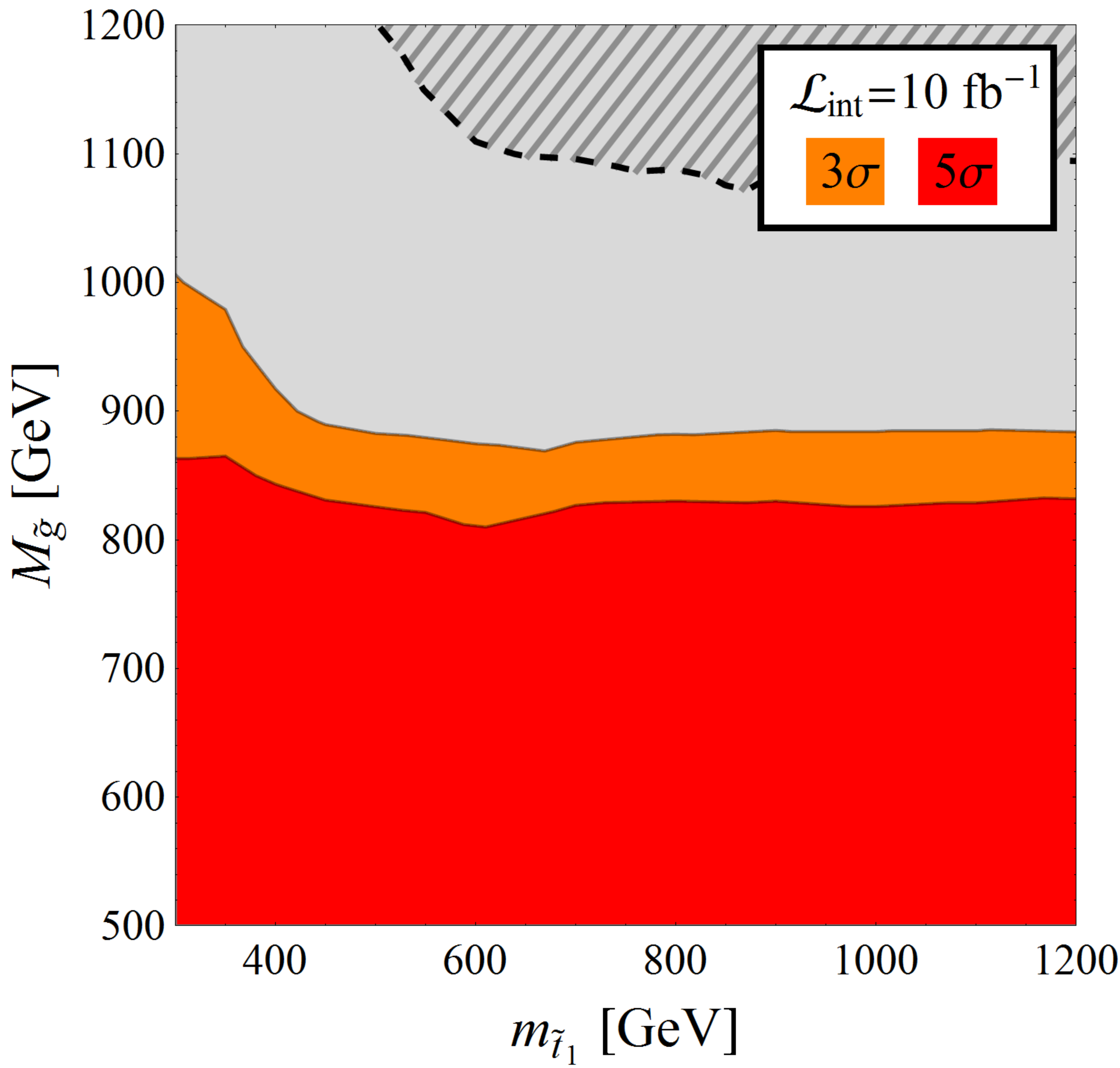}
  \includegraphics*[width=0.45\columnwidth]{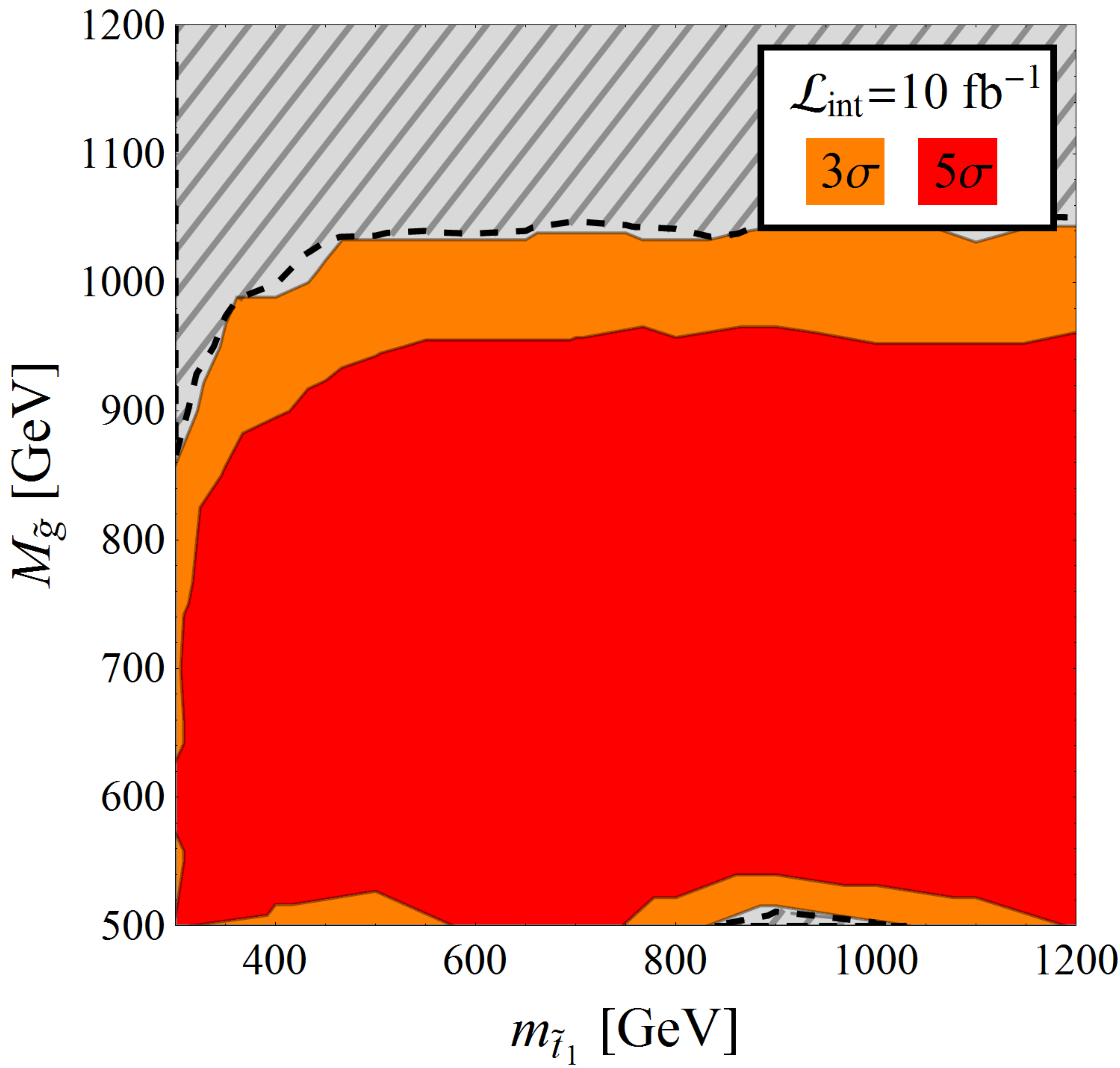} }
\caption{
Signal significance contours (orange =$3\sigma$, red=$5\sigma$) in the $(m_{\tilde t_1},M_{\tilde g})$ plane
with $\mathcal{L}_{\mathrm{int}} = 10~\mathrm{fb}^{-1}$ at the $\sqrt{s}=7$~TeV LHC.  For
all panels, events must satisfy the precuts.  Additional cuts which can be applied are
$\met > 300~\gev$ (upper right), $N_j \geq 8$ (lower left), and $\met > 300~\gev$, $N_j \geq 8$
(lower right).
In the dark gray hashed region bounded by the dashed contour, less than 5 signal events are expected.
\label{fig:SignificancePanels}}
\end{figure}

The general features of these cuts are evident from figure~\ref{fig:SignificancePanels}.
In regions of parameter space where one can pair-produce particles whose decays yield
more than two tops (i.e., when gluinos are accessible to the LHC), the elevated
jet cut improves signal significance.  In regions of parameter space where the decays
of new particles yield energy in $\met$ ($M_{\tilde g} \gg m_{\tilde t_1} \gg M_{\tilde N_1}$),
an elevated $\met$ is effective in distinguishing signal from background.

One can compare the detection prospects for this search strategy with those of other
strategies~\cite{TopsPlusMET,Kane:2011zd}.
For example, a search for gluino pair-production using the channel with 1 charged lepton and
4 $b$-tagged jets was estimated to produce a $5\sigma$-sensitivity for  $M_{\tilde g}$ as large
as $650~\gev$ with $1~\ifb$ of data at $7~\tev$~\cite{Kane:2011zd}.  For the same luminosity and
center-of-mass energy, the search strategy discussed here is estimated to provide $5\sigma$-sensitivity for
$M_{\tilde g}$ as large as $720~\gev$, thus indicating that this search strategy can be at least
comparable to other strategies.

For an analysis involving a broad distribution of jets, one must worry about
systematic uncertainties in the estimation of the background.  This concern can be
addressed by requiring that the ratio of signal events ($S$) to background events ($B$) be
large enough.  For all of the elevated cuts imposed in figure~\ref{fig:SignificancePanels}, we
find $S/B > 0.1$.  However, this constraint is not satisfied for the light stop mass region in
the case where only the precuts are imposed.  To control systematic uncertainties, it thus may
be desirable to impose an additional cut to reduce background.

Our choices of elevated cuts ($N_j \geq 8$ and/or $\met > 300~\gev$)
are optimized for a 7 TeV $10~\ifb$ LHC run.  For gluino pair-production, an even higher jet
cut will improve $S/B$, but will reduce the number of signal events dramatically.  For an LHC run
with higher luminosity, a more elevated jet cut may be desirable.  Similarly, for a run at higher
energy, a more elevated $\met$ cut may be desirable.

\section{Conclusion}

Although this search strategy has been applied to the MSSM with
minimal flavor violation  and a particular mass spectrum (the only light sparticles are
the gluino, lightest stop and lightest neutralino), its applicability is
more general.  This strategy can be effective for any model in which one
can pair-produce new particles which decay primarily to tops and dark matter.  The
signature is many jets, $\met$, and no charged leptons.  The charged
lepton veto removes SM background events with $\met$ arising from $W \rightarrow l\nu$.
An elevated cut on the number of jets can improve sensitivity to models which produce
multiple tops from the decay of new heavy particles, while an elevated $\met$ cut can improve
sensitivity to models wherein the kinematics of the decay chain favor the production of
boosted dark matter.  Searches for this signature have been implemented~\cite{JetsPluMETExp},
with results which corroborate our expectations.
This strategy may be effective in natural SUSY searches.

{\bf Acknowledgements}

We are grateful to J.~Bramante, X.~Tata, and B.~Thomas, and to the organizers
of ICHEP2012.  This work is
supported in part by Department of Energy grant DE-FG02-04ER41291.

\end{document}